\documentclass[prd,onecolumn,nofootinbib,aps,tightenlines,preprintnumbers,notitlepage,longbibliography,superscriptaddress,12pt]{revtex4-1}

\usepackage[dvipsnames]{xcolor}
\usepackage{color}
\usepackage{tikz}
\usepackage[caption=false]{subfig}
\usepackage{hyperref}
\usepackage{multirow}
\usepackage{siunitx}
\usepackage[export]{adjustbox}
\usepackage{graphicx}
\usepackage{dcolumn}
\usepackage{bm}
\usepackage[normalem]{ulem}
\usepackage{epstopdf}
\usepackage{tabularx}
\usepackage{suffix}
\usepackage{mathtools}
\usepackage{aas_macros}

\graphicspath{ {figures/} }
\usetikzlibrary{arrows.meta,
                chains,
                positioning,
                quotes,
                shapes.geometric}

\tikzset{FlowChart/.style={
startstop/.style = {rectangle, rounded corners, draw, fill=blue!10,
                    minimum width=3cm, minimum height=1cm, align=center,
                    on chain, join=by arrow},
  process/.style = {rectangle, draw, fill=orange!10,
                    minimum width=3cm, minimum height=1cm, align=center,
                    on chain, join=by arrow},
 decision/.style = {diamond, aspect=1.5, draw, fill=green!30,
                    minimum width=3cm, minimum height=1cm, align=center,
                    on chain, join=by arrow},
       io/.style = {trapezium, trapezium stretches body,   
                    trapezium left angle=70, trapezium right angle=110,
                    draw, fill=blue!30,
                    minimum width=3cm, minimum height=1cm,
                    text width =\pgfkeysvalueof{/pgf/minimum width}-2*\pgfkeysvalueof{/pgf/inner xsep},
                    align=center,
                    on chain, join=by arrow},
    arrow/.style = {thick,-Triangle}
                        }
        }

\newcolumntype{C}{>{\centering\arraybackslash}X}
\newcolumntype{R}{>{\raggedleft\arraybackslash}X}

\newcommand{\be}{\begin{eqnarray}}

\newcommand{\ee}{\end{eqnarray}}

\definecolor{colorA}{HTML}{9467bd}
\definecolor{colorB}{HTML}{bcbd22}
\definecolor{colorC}{HTML}{2ca02c}

\newcommand{\smu}{Department of Physics,
Southern Methodist University, 3215 Daniel Ave, Dallas, TX 75275, USA}

\begin{document}

\title{Insights for Early Dark Energy with Big Bang Nucleosynthesis}

\author{Christopher Cook}
\affiliation{\smu}

\author{Joel~Meyers}
\affiliation{\smu}

\date{\today}

\begin{abstract}

Big Bang Nucleosynthesis (BBN), as one of the earliest processes in the universe accessible to direct observation, offers a powerful and independent probe of the cosmic expansion history. With recent advances in both theory and observation, including efficient and flexible BBN codes, percent-level measurements of primordial deuterium and helium-4 abundances, refined measurements of nuclear reaction rates, and precise determinations of the baryon density from the cosmic microwave background, particularly keen insights can be gained from BBN. In this work, we leverage these developments to place model-independent constraints on deviations from the Standard Model expansion history during BBN. Using the latest abundance data, we apply principal component analysis to identify the most constrained and physically meaningful modes of expansion history variation. This approach allows us to impose the most general constraints on early dark energy during the epoch of BBN. We further examine whether general modifications to the expansion rate could alleviate the long-standing lithium problem. Our results demonstrate that BBN, sharpened by modern data and statistical techniques, remains an indispensable probe of dark energy and new physics in the early universe.

\end{abstract}

\maketitle

\section{Introduction}
\label{sec:Introduction}

Big Bang Nucleosynthesis (BBN) is a cornerstone of modern cosmology. As the earliest process in the universe that can currently be tested against direct observations, BBN provides one of our most powerful windows into the physics of the early universe. 
There exists a well-established theoretical framework connecting the light element abundances to fundamental cosmological parameters; see Refs.~\cite{Cyburt:2015mya,Fields:2019pfx,Grohs:2023voo,ParticleDataGroup:2024cfk} for recent reviews. 
The predictions of standard BBN for the primordial abundances of the light elements, particularly helium-4 and deuterium, are in remarkable agreement with measurements, confirming the robustness of the Standard Model of particle physics and cosmology~\cite{ParticleDataGroup:2024cfk}. At the same time, because the abundance of these elements depends sensitively on the strength of fundamental interactions and on the cosmic history, BBN offers a unique probe of new physics in the early universe.

Recent years have produced rapid progress in both observational and theoretical aspects of BBN. Measurements of the primordial abundances of helium-4~\cite{Aver:2020fon,Valerdi:2019beb,Fernandez:2019hds,Kurichin:2021ppm,Hsyu:2020uqb,Valerdi:2021abc,Aver:2021rwi,ParticleDataGroup:2024cfk} and deuterium~\cite{Cooke:2013cba, Cooke:2016rky, Riemer-Sorensen:2014aoa, Riemer-Sorensen:2017pey,Balashev:2015hoe, Zavarygin:2017cov, Cooke:2017cwo, ParticleDataGroup:2024cfk} have reached unprecedented precision, yielding abundances consistent with Standard Model predictions at the percent level. Experimental efforts, such as those by the LUNA collaboration~\cite{Mossa:2020gjc}, have reduced uncertainties in the nuclear reaction rates that govern BBN yields. Moreover, observations of the cosmic microwave background (CMB) now tightly constrain the baryon density~\cite{Planck:2018vyg,AtacamaCosmologyTelescope:2025blo,SPT-3G:2025bzu}, eliminating one of the largest sources of uncertainty in comparing BBN theory with data.  Numerical computation of the predictions of BBN for a range of cosmological scenarios can be achieved through various publicly available software packages~\cite{Arbey:2011nf,Arbey:2018zfh, Pisanti:2007hk,Consiglio:2017pot,Pitrou:2019nub,Burns:2023sgx,Giovanetti:2024zce}. Together, these advances establish BBN as an incisive tool for testing physics beyond the Standard Model.

One motivation for such tests arises from the ongoing Hubble tension, the discrepancy between the Hubble constant inferred from local distance ladder measurements and that inferred from CMB observations within the $\Lambda$CDM framework~\cite{DiValentino:2021izs, Kamionkowski:2022pkx}. Early dark energy (EDE), in which an additional energy component contributes non-negligibly to the total energy density before recombination, has emerged as a compelling candidate for resolving this tension~\cite{Kamionkowski:2022pkx}. While CMB observations from collaborations like Planck~\citep{Planck:2018vyg}, ACT~\citep{AtacamaCosmologyTelescope:2025blo, AtacamaCosmologyTelescope:2025nti}, and SPT~\cite{SPT-3G:2025bzu,SPT-3G:2025vyw} place strong bounds on such scenarios, complementary probes of the expansion history are crucial for testing the viability of EDE and other early-universe modifications.

Several avenues have been explored in this direction~\cite{Allahverdi:2020bys}. For example, small-scale structure places constraints on modifications to the expansion history at early times~\cite{Erickcek:2011us,Delos:2018ueo}, while gravitational wave backgrounds~\cite{Caprini:2018mtu} are, in principle, sensitive to the equation of state during very early epochs~\cite{Watanabe:2006qe,Barenboim:2016mjm}. Studies of the effects of the pre-recombination expansion history using principal component methods~\citep{Samsing:2012qx} and constraints from the matter power spectrum on early dark energy~\citep{Sobotka:2024ixo} have provided complementary insights into possible deviations from the standard expansion history. Yet BBN offers a particularly powerful and direct probe, since light element abundance yields are directly sensitive to the expansion rate during the first few minutes after the Big Bang.

Previous studies have shown that BBN abundances can be used to constrain the effective equation of state during nucleosynthesis~\cite{Carroll:2001bv,Arbey:2009gt,McKeen:2024voa}, but the steadily improving precision of abundance measurements now makes it possible to go further. Recently, Ref.~\citep{An:2023buh} explored a model-independent reconstruction of the expansion and thermal histories constrained by the light element abundances. In this paper, we use the latest abundance data to derive general constraints on the expansion history during BBN. However, unlike Ref.~\citep{An:2023buh}, we employ principal component analysis to extract the most robust, model-independent features of these constraints, allowing us to place limits on broad classes of early-universe scenarios, including those motivated by early dark energy. In addition, we revisit the long-standing lithium problem, the persistent discrepancy between predicted and observed primordial lithium abundances~\cite{Fields:2011zzb,ParticleDataGroup:2024cfk}. While it is well established that uncertainties in nuclear reaction rates cannot resolve this discrepancy~\cite{ParticleDataGroup:2024cfk}, we investigate whether a general modification to the expansion history could mitigate the tension.

The remainder of this paper is organized as follows. In Section~\ref{sec:BBN}, we review the current observational status of Big Bang Nucleosynthesis and summarize the relevant data. Section~\ref{sec:Methodology} describes our methodology, including the BBN calculations, the construction of a basis for the Fisher analysis, and the subsequent Fisher matrix, principal component, and Markov chain Monte Carlo (MCMC) analyses. In Section~\ref{sec:Results}, we present our results, followed by a discussion of their implications and concluding remarks in Section~\ref{sec:Conclusion}.

\section{Big Bang Nucleosynthesis}
\label{sec:BBN}

Big Bang Nucleosynthesis refers to the formation of light nuclei, primarily hydrogen, deuterium, helium-3, helium-4, and lithium-7, with trace amounts of heavier nuclei, within the first few minutes after the Big Bang, when the Universe cooled sufficiently for nuclear reactions to proceed. Governed by the Standard Model of particle physics and the well-understood thermodynamics of the expanding early Universe, BBN serves as a direct and useful probe of cosmology at the earliest epoch to which we currently have observational access. The predicted primordial abundances are sensitive to several key parameters, including the baryon-to-photon ratio ($\eta$), the light relic density, the strengths of interactions among particles, the neutron lifetime, and the expansion history of the early Universe.

As such, BBN plays a pivotal role in constraining both standard cosmological parameters and potential new physics beyond the Standard Model, such as decaying particles, asymmetric reheating, or varying fundamental constants~\cite{ParticleDataGroup:2024cfk}. Among the most robust predictions are those for deuterium and helium-4, whose abundances can now be measured with high precision in nearly pristine astrophysical environments. The concordance between BBN predictions and observational data for these elements offers strong support for the standard cosmological model. However, the observed primordial abundance of lithium-7 remains significantly lower than theoretical expectations, a discrepancy known as the ``lithium problem," which may point to underestimated observational errors or hints of new physics~\cite{Fields:2011zzb,Fields:2019pfx,ParticleDataGroup:2024cfk}.

\subsection{Observational Status}
\label{subsec:BBN_Observations}

Recent observations of primordial light element abundances continue to support the standard cosmological model for most isotopes while highlighting persistent anomalies. We use the Particle Data Group (PDG)~\cite{ParticleDataGroup:2024cfk} recommended values of observed abundance ratios throughout this work, which are summarized in Table~\ref{tab:abundances}. Deuterium remains the most accurately measured primordial isotope, with high-redshift quasar absorption systems yielding D/H = $(2.547 \pm 0.029) \times 10^{-5}$,
in excellent agreement with BBN predictions when using the CMB-derived baryon density.  
Similarly, helium-4 determinations from extragalactic HII regions, aided by improved emissivity corrections and infrared line data, support a mass fraction $Y_p = 0.245 \pm 0.003$, matching theoretical expectations. The primordial abundance of helium-3, however, remains poorly constrained due to its detection being limited to metal-rich Galactic environments. Lithium-6, once thought to be overabundant, is now strongly constrained in both Galactic halo stars and the Small Magellanic Cloud (SMC), with no confirmed detections, effectively ruling out a primordial $^6$Li plateau and supporting the case for stellar depletion.

Lithium-7 remains a significant challenge to the standard BBN paradigm. The Spite plateau in metal-poor halo stars yields a $^7$Li/H ratio of $(1.6 \pm 0.3) \times 10^{-10}$, a factor of three below the BBN+CMB prediction of $(4.72 \pm 0.7) \times 10^{-10}$. New observations reveal dispersion and decreasing $^7$Li levels at metallicities [Fe/H] $\lesssim -2.7$, pointing to stellar depletion as a plausible explanation~\cite{Fields:2022mpw}. In parallel, stringent upper limits on $^6$Li/$^7$Li, such as $< 0.007$ in HD 84937 and $< 0.1$ in the SMC’s Sk 143 sightline, suggest significant $^6$Li depletion. Since $^6$Li is more fragile than $^7$Li, these findings imply that comparable $^7$Li destruction is astrophysically feasible, potentially resolving the discrepancy without requiring new physics.

Yet, extragalactic evidence challenges this resolution. The detection of gas-phase $^7$Li in the SMC with an abundance $A(^7\mathrm{Li}) \approx 2.2$, unaffected by stellar processes, still falls below BBN expectations~\cite{Molaro:2024wxa}. This implies a primordial shortfall that stellar depletion alone cannot explain. While Galactic $^6$Li non-detections support an astrophysical solution, the SMC results suggest a possible need for physics beyond the Standard Model. Although stellar depletion offers a compelling explanation within the Milky Way, the lithium problem may ultimately point to deeper cosmological questions.

\begin{table}[h!]
    \centering
    \begin{tabular}{lcc}
    \hline
    \textbf{Element} & \textbf{\hspace{0.5cm} Observed Abundance \hspace{0.5cm} } & \textbf{Standard BBN Prediction} \\
    \hline
    Deuterium (D/H) & $(2.547 \pm 0.029) \times 10^{-5}$ & $(2.459 \pm 0.036) \times 10^{-5}$ \\
    Helium-4 (mass fraction $Y_p$) & $0.245 \pm 0.003$ & $0.2471 \pm 0.0002$ \\
    Lithium-7 (Li/H) & $(1.6 \pm 0.3) \times 10^{-10}$ & $(4.72 \pm 0.7) \times 10^{-10}$ \\
    \hline
    \end{tabular}
    \caption{Primordial light element abundances: observations vs. Standard BBN predictions (assuming $\eta_{10} = 6.143 \pm 0.190$). This work uses the PDG~\cite{ParticleDataGroup:2024cfk} recommended values of observed abundance ratios.
    }
    \label{tab:abundances}
\end{table}

\section{Methodology}
\label{sec:Methodology}

\subsection{BBN Calculations}
\label{subsec:BBN_Calc}

To study the effects of a non-standard expansion history on BBN, we use a modified version of \texttt{AlterBBN}~\citep{Arbey:2011nf, Arbey:2018zfh}. This publicly available code numerically solves the coupled differential equations governing the evolution of light nuclei in the early universe. Boltzmann solvers like \texttt{AlterBBN} are essential tools in cosmology because they compute the time evolution of particle species based on their interactions, decay rates, and the universe's expansion history. Specifically for BBN, \texttt{AlterBBN} integrates the nuclear reaction network for light elements, such as deuterium, helium-3, helium-4, and lithium-7, while simultaneously solving the Friedmann equation that determines the expansion rate as a function of time $H(t)$.

\begin{figure}[h!] 
    \centering
    \includegraphics[width=1.0\textwidth]{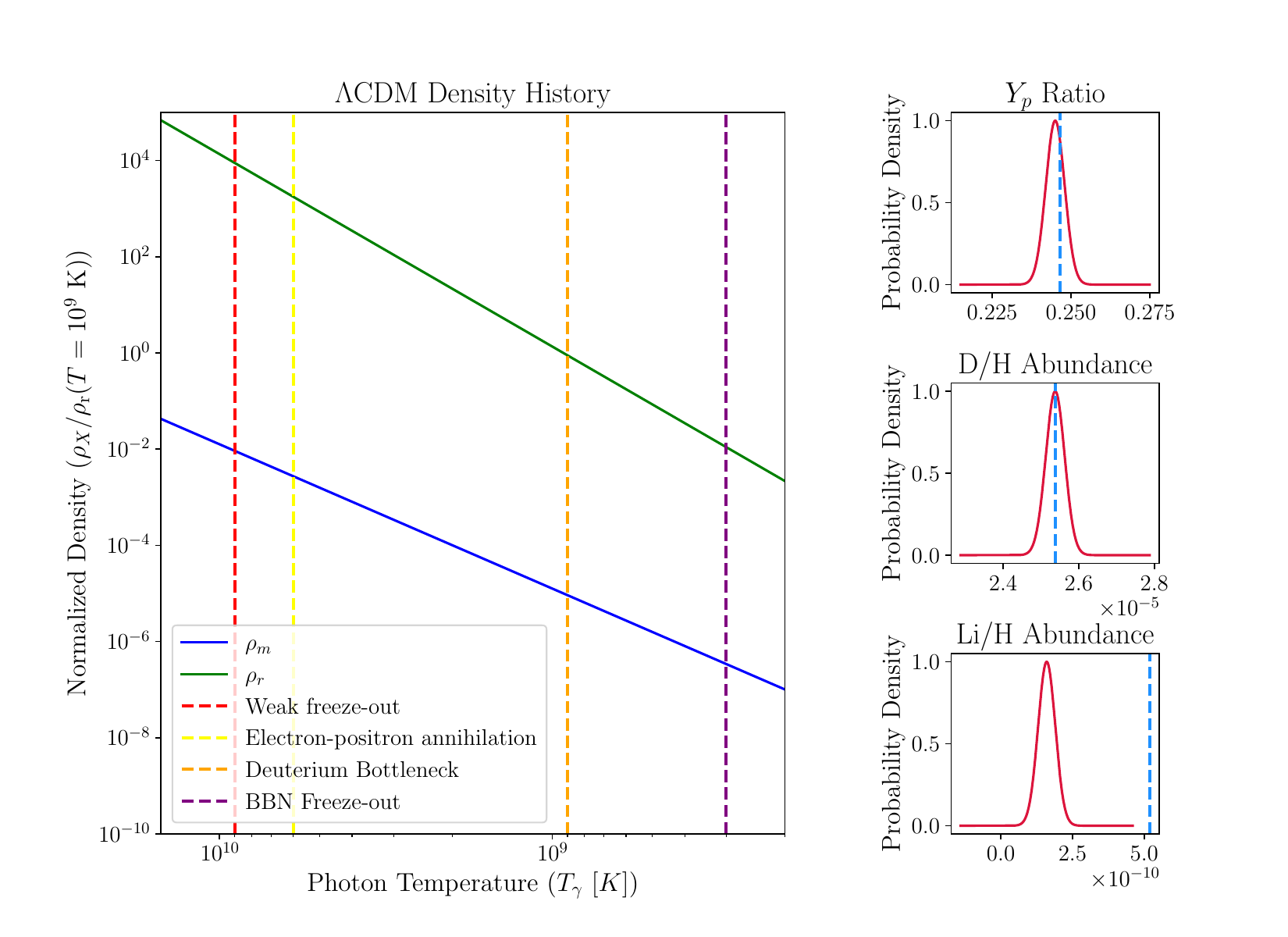} 
    \caption{Evolution of energy densities in the standard cosmological history, highlighting various temperature scales of physical relevance to BBN.  The right panels show the predictions of the primordial abundances derived from Standard BBN assuming $\eta_{10} = 6.143 \pm 0.190$ (blue dashed lines) compared to the observational constraints (red solid lines). 
    }
    \label{fig:LCDM}
\end{figure}

In the standard cosmological scenario, the energy content during BBN is dominated by radiation, with the Hubble rate evolving accordingly. \texttt{AlterBBN} computes light element abundances based on these inputs, under the assumption of a \(\Lambda\)CDM background.  Figure~\ref{fig:LCDM} shows the evolution of the radiation and matter densities during the epoch of BBN in the standard cosmological model, along with the predicted light element abundances compared to their observed values. Moving beyond the standard scenario, the expansion rate is sensitive to additional energy components, such as EDE, especially at early times. To incorporate the effects of additional contributions to the energy density of the early Universe, we extend \texttt{AlterBBN} to accept a user-specified table of temperatures and corresponding dark energy densities, which modifies the total energy density and alters the Hubble expansion rate accordingly. This extension allows us to probe arbitrary EDE histories beyond the limitations of analytic or parameterized models.  We will refer to the additional energy density as EDE, though the constraints we derive apply generally to the presence of any form of additional energy density present during the BBN epoch, so long as it does not interact non-gravitationally with the Standard Model degrees of freedom.

The modified code computes the primordial abundances for each dark energy history, directly mapping EDE behavior and observable light element yields. This allows us to test at what temperatures the light elements are most sensitive to changes in the density of early dark energy. 

\subsection{Constructing a Basis for Fisher Analysis}
\label{subsec:Basis}

\begin{figure}[h!] 
    \centering
    \includegraphics[width=1.0\textwidth]{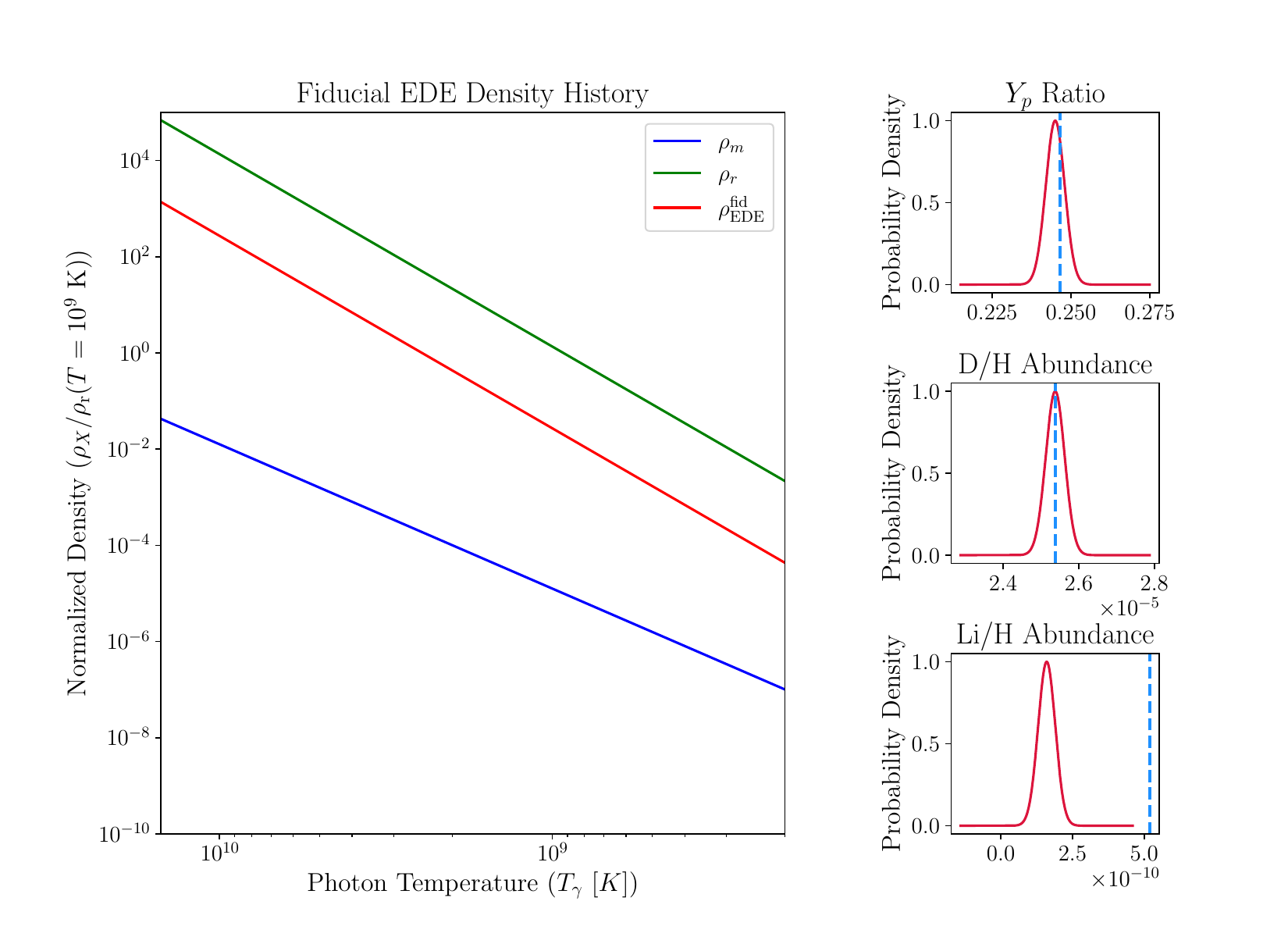} 
    \caption{Evolution of densities in the fiducial early dark energy model that we perturb to generate the basis of our Fisher Information Matrix.  The EDE density in the fiducial model (red line) is chosen to behave like radiation with a density smaller than the Standard Model radiation (green line) by a factor of 50.  The right panels demonstrate that the predicted abundances in this fiducial model are essentially identical to that of Standard BBN.
    }
    \label{fig:FidEDE}
\end{figure}

To evaluate the sensitivity of BBN to arbitrary EDE histories, we construct a localized basis by perturbing a fiducial model of dark energy density evolution. This was done to most effectively explore the relevant models of EDE while identifying the temperature range at which the light elements are most sensitive to change. The fiducial model follows a standard radiation-like dilution, $\rho_{\mathrm{EDE}}^\mathrm{fid}(T) \propto T^4$, which ensures that the energy density remains subdominant at all times while tracking the behavior of radiation. We set the fiducial model energy density to be a factor of 50 smaller than that of the Standard Model radiation; see Figure~\ref{fig:FidEDE}.

We discretize the temperature range relevant for BBN, from $T \sim 10$~MeV to $T \sim 0.001$~MeV, into 150 logarithmically spaced bins. For each bin, we generate a perturbation by scaling the energy density $\rho_{\mathrm{EDE}}(T_i)$ in a single bin by a small factor $1 + \epsilon$ (e.g., $\epsilon = 0.01$), keeping all other bins fixed. This results in a set of perturbed energy density tables $\{\rho^{(i)}_{\mathrm{EDE}}(T)\}$ which serve as a basis of perturbations about the fiducial model.

Each perturbed table is passed into the modified \texttt{AlterBBN} code to compute the corresponding light element abundances. By finite differencing the abundance shifts $\Delta Y_a$ with respect to the perturbations $\Delta \rho_i = \epsilon\rho_\mathrm{EDE}^\mathrm{fid}(T_i)$, we numerically evaluate the derivatives:
\begin{equation}
\frac{\partial Y_a}{\partial \rho_i} \approx \frac{Y_a^{(i)} - Y_a^{(0)}}{\Delta \rho_i},
\label{derivative}
\end{equation}
where $Y_a^{(0)}$ is the abundance from the fiducial model, and $Y_a^{(i)}$ is the abundance resulting from a perturbation in bin $i$. These derivatives are used to compute the Fisher matrix.

\subsection{Fisher Matrix and Principal Component Analysis}

The Fisher matrix quantifies how precisely the EDE density can be constrained in each temperature bin, based on the response of observable abundances and their measurement uncertainties. It is defined as:
\begin{equation}
F_{ij} = \sum_{a} \frac{1}{\sigma_a^2} \left( \frac{\partial Y_a}{\partial \rho_i} \right) \left( \frac{\partial Y_a}{\partial \rho_j} \right),
\label{fisherMatrix}
\end{equation}
where $\sigma_a$ is the uncertainty in the observational determination of the light element $Y_a$ (e.g., D/H, $Y_p$, Li/H).  Of course, with only a small number of observed primordial abundances, it is not possible to independently constrain the expansion history in every temperature bin. We therefore perform a principal component analysis (PCA) by diagonalizing the Fisher matrix:
\begin{equation}
    F = W^\top \Lambda W,
    \label{PCA}
\end{equation}
where $\Lambda$ is the diagonal matrix of eigenvalues and $W$ is the matrix of orthonormal eigenvectors. Each eigenvector represents an eigenmode or principal component of the EDE history to which BBN is sensitive, organized in descending order of sensitivity. The corresponding eigenvalues indicate the degree of constraint on each mode, with large eigenvalues corresponding to directions in parameter space that are well-constrained by current data. Since we have 3 well-constrained observables (the deuterium, helium-4, and lithium abundances), the Fisher matrix is a rank 3 matrix and thus has 3 well-constrained eigenmodes. 

The eigenmode amplitudes \( \alpha_i \) are used to determine the dark energy density history as a perturbation around a fiducial radiation-like profile:
\begin{equation}
    \log_{10}(\rho_{\mathrm{EDE}}(T)) = \log_{10}(\rho_\mathrm{EDE}^\mathrm{fid}(T)) + \sum_{i=1}^{3} \log_{10}\left( \alpha_i e_i(T)\right),
    \label{ModelCalculation}
\end{equation}
where \( e_i(T) \) are the eigenmodes derived from the Fisher matrix analysis, and \( \rho_\mathrm{EDE}^{\mathrm{fid}}(T) \propto T^4 \).

\begin{figure}[h!] 
    \centering
    \includegraphics[width=1.0\textwidth]{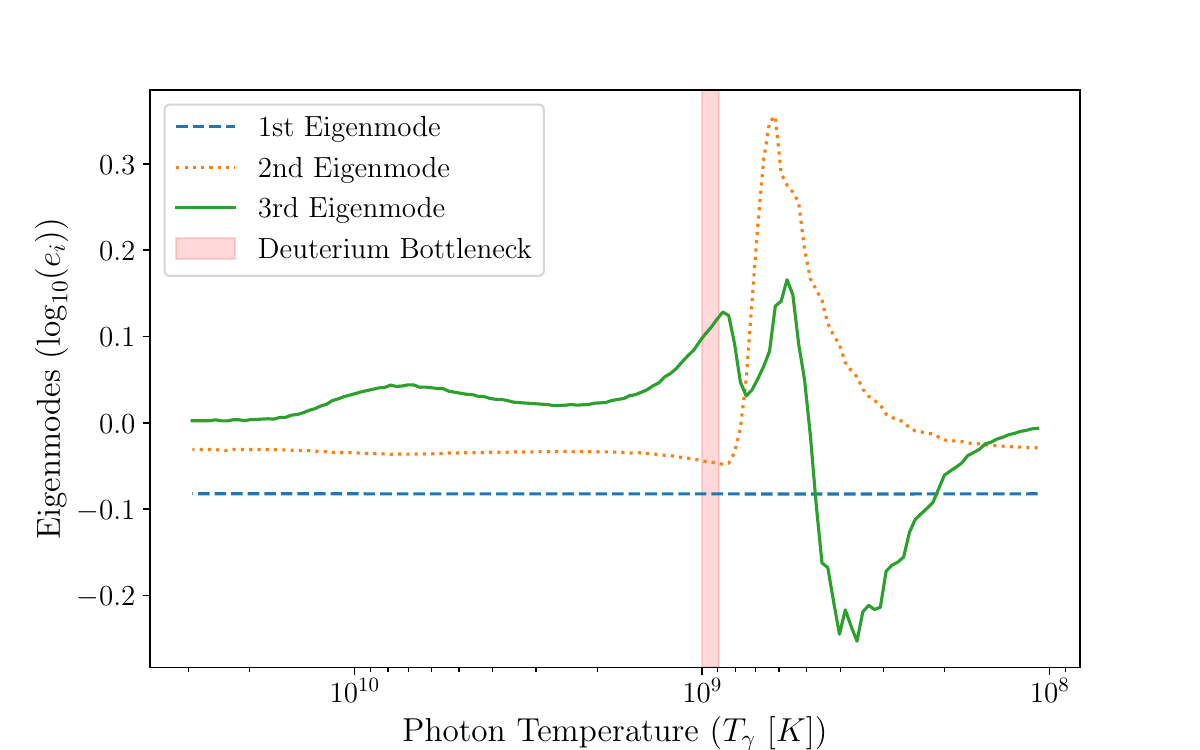} 
    \caption{First three principal components of the dark energy density (defined here as a deviation from radiation-like behavior) to which primordial abundances are sensitive. The vertical axis shows the difference from the fiducial, radiation-like, energy density in log-space. The horizontal axis shows the corresponding temperature in log-space. The first principal component represents a uniform shift up or down in log-log space. The second and third eigenmodes represent an injection of energy during and after the Deuterium Bottleneck that occurs when the photon temperature is around $10^9$~K. }
    \label{fig:Eigenstates}
\end{figure}

These principal components provide a powerful basis for model-independent constraints: any arbitrary EDE history can be projected onto the eigenbasis, allowing one to reconstruct the constrained and unconstrained features of the model in terms of measurable quantities. The plot of the three eigenmodes are shown in Figure~\ref{fig:Eigenstates}.  
The most strongly constrained mode is degenerate with a uniform increase to the radiation density.  However, as seen in the second and third eigenmodes, the primordial abundances are also sensitive to temperature-dependent changes in the expansion history, especially those that affect expansion in the temperature range $10^9~\mathrm{K}>T_\gamma>3\times10^8$~K.  Qualitatively similar conclusions were drawn from the binned expansion history analysis presented in Ref.~\cite{An:2023buh}, where the authors also found that modifications to the expansion rate around this temperature range have the most significant impact on light element abundances. The agreement between the PCA-based and binned analyses provides a strong validation that the constraints identified here are robust to the choice of parametrization.

\subsection{Markov Chain Monte Carlo Analysis}
\label{subsec:MCMC}

To explore the full posterior distribution of EDE histories and relevant cosmological parameters, we perform a Markov Chain Monte Carlo (MCMC) analysis using the \texttt{Cobaya} framework~\cite{Torrado:2020dgo}. \texttt{Cobaya} (Code for Bayesian Analysis) is a modular, extensible Python-based platform for efficient Bayesian inference, particularly well suited to hierarchical physical models with complex interdependencies.

Our parameter space includes both standard BBN inputs and early dark energy features:
\begin{itemize}
    \item \( \eta_{10} \equiv 10^{10} \times \frac{n_b}{n_\gamma} \): the baryon-to-photon ratio, with a Gaussian prior $\eta_{10} = 6.104 \pm 0.058$, consistent with the constraint derived from CMB observations~\cite{Planck:2018vyg,ParticleDataGroup:2024cfk},
    \item \( \tau_n \): the neutron lifetime, with a Gaussian prior $\tau_{n} = 878.4 \pm 0.5$~s, consistent with laboratory measurements~\cite{ParticleDataGroup:2024cfk},
    \item \( \alpha_1, \alpha_2, \alpha_3 \): amplitudes of the first three PCA eigenmodes of the EDE density, each with a flat prior [-20,20].
\end{itemize}

Each set of parameters defines a complete model of early dark energy. Note that the first principal component of the dark energy density is degenerate with that of the radiation density, typically parameterized by $N_\mathrm{eff}$, so we did not separately vary $N_\mathrm{eff}$ in our analysis.  Several previous studies have shown how observations of primordial abundances can be used to constrain the light relic density~\cite{ParticleDataGroup:2024cfk,Cyburt:2015mya,Fields:2019pfx,Lague:2019yvs,Dvorkin:2022jyg,Yeh:2022heq}.

This pipeline is managed in \texttt{Cobaya} by linking external theory components (in our case, a Python wrapper to a modified version of \texttt{AlterBBN}) with a custom likelihood module that computes the Gaussian likelihood of predicted light element abundances:
\begin{equation}
\ln \mathcal{L} = -\frac{1}{2} \sum_a \left( \frac{Y_a^{\mathrm{model}} - Y_a^{\mathrm{obs}}}{\sigma_a} \right)^2,
\label{Likelihood}
\end{equation}
with \( a \in \{\mathrm{D/H}, Y_p, \mathrm{Li/H}\} \) and \( \sigma_a \) denoting their respective observational uncertainties.

Sampling is performed using \texttt{Cobaya}’s built-in Metropolis-Hastings algorithm, which automatically groups parameters into fast and slow blocks based on the computational cost of the components that depend on them~\cite{Lewis:2002ah,Lewis:2013hha,Neal:2005uqf,Torrado:2020dgo}. This automatic parameter blocking and caching of intermediate results greatly improves sampling efficiency. Chain convergence is monitored using the Gelman-Rubin \( R - 1 \) statistic, with a typical stopping criterion of \( R - 1 < 0.01 \).

Posterior samples are processed using \texttt{GetDist}~\cite{Lewis:2019xzd}, which interfaces seamlessly with \texttt{Cobaya}'s output, providing marginal distributions, credible intervals, and covariance matrices. In particular, we analyze the posterior distributions of the eigenmode amplitudes \( \alpha_i \), and project the results back into temperature space to reconstruct allowed EDE histories and identify epochs most sensitive to BBN constraints. We use the observational constraints on primordial abundances from the PDG~\cite{ParticleDataGroup:2024cfk}, summarized above in Table~\ref{tab:abundances}.

\section{Results}
\label{sec:Results}

\begin{figure}[htbp] 
    \centering
    \includegraphics[width=0.7\textwidth]{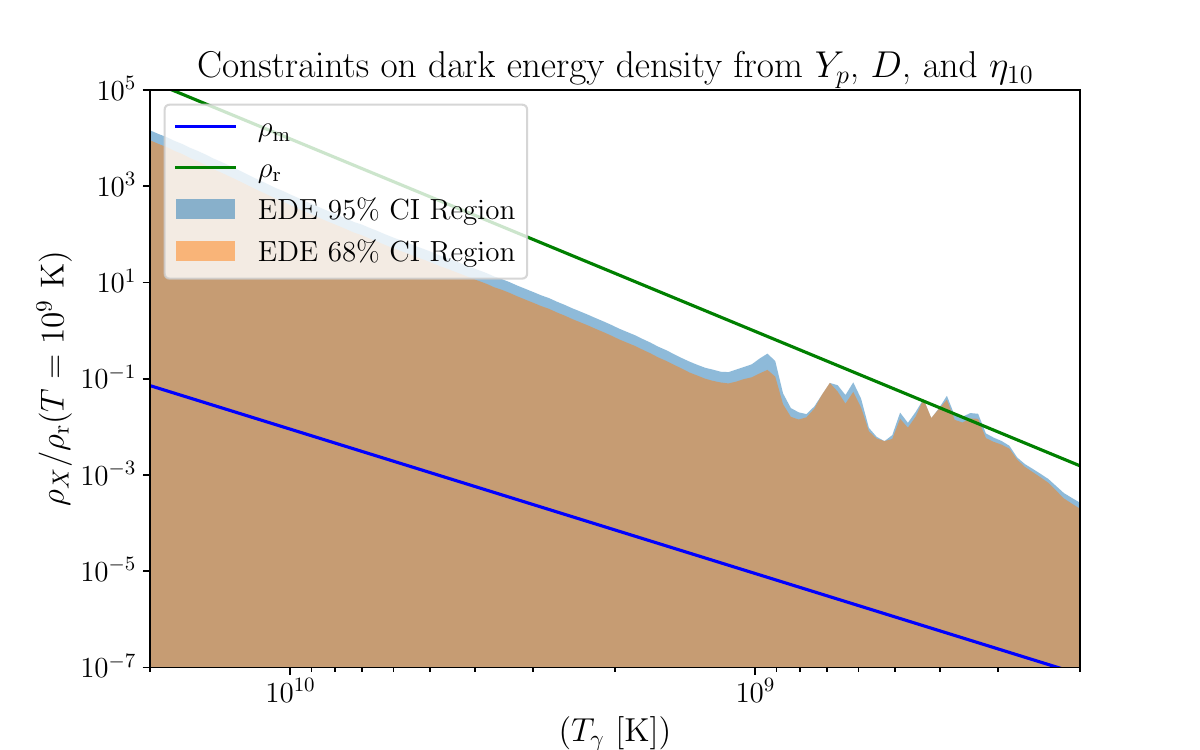} 
    \caption{Constraints on EDE using observed values of $Y_{p}$, $D$, and $\eta_{10}$. Shown are the matter and radiation densities, together with the 68\% and 95\% confidence intervals for EDE.}
    \label{fig:NoCMB}
\end{figure}

Figure~\ref{fig:NoCMB} shows the allowed EDE histories consistent with the observed primordial helium-4 and deuterium abundances. 
In this scenario, the EDE component is generally constrained to remain subdominant throughout the epoch of BBN, except in the narrow temperature range $3\times10^{8}\,\mathrm{K} \lesssim T \lesssim 5\times10^{8}\,\mathrm{K}$, where  constraints are weaker, and an EDE component with energy density comparable to the radiation density is permitted. A  second poorly constrained region is also visible around or just after the deuterium bottleneck, indicating a localized allowance for additional EDE energy density during that stage. It should be noted, however, that a localized spike in energy density is unlikely to be realized in a self-consistent physical model of EDE.  For example, the equation of state of any fluid obeying the null energy condition satisfies $w\geq-1$, and thus has an energy density that either remains constant or dilutes with the expansion of space.  Furthermore, a free scalar field has $w=1$ with an energy density scaling as $\rho\propto a^{-6}$, while the presence of a non-vanishing potential would lead to a slower decrease in energy density.  Thus the bounds we present here should be interpreted as indicating the regions where observational constraints are more or less constraining, while bearing in mind that it may not be possible to saturate the upper limits in the less well-constrained regions in a self-consistent model.

These results are broadly consistent with the binned reconstruction of the expansion history presented in Ref.~\cite{An:2023buh}, confirming that the light-element abundances primarily constrain deviations in the Hubble rate near the onset of nucleosynthesis. The PCA framework developed here extends that earlier analysis by providing a continuous, model-independent mapping of the allowed expansion histories, revealing not only where the expansion rate is tightly constrained but also which temperature ranges remain weakly bounded by current data.

\begin{figure}[htbp] 
    \centering
    \includegraphics[width=0.7\textwidth]{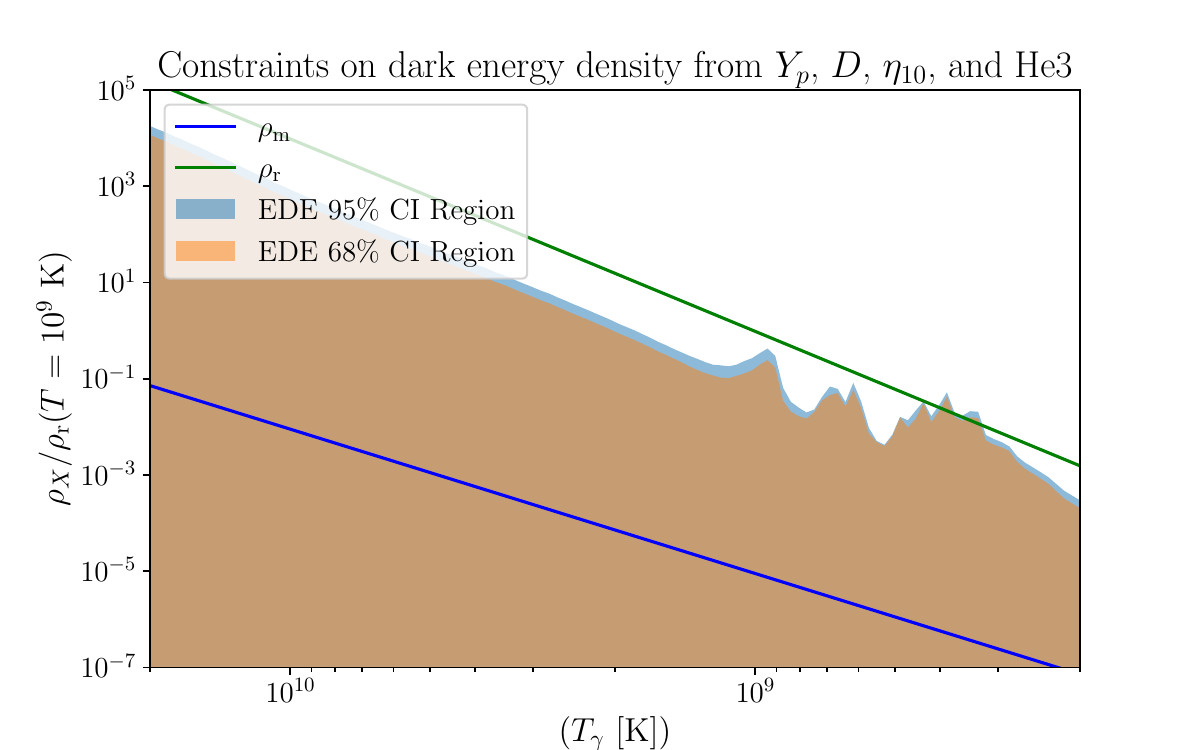} 
    \caption{Constraints on EDE from observed values $Y_{p}$, $D$, $\eta_{10}$, and a forecasted 1\% constraint on the helium-3 abundance. Shown are the matter and radiation densities, together with the 68\% and 95\% confidence intervals for EDE.}
    \label{fig:NoHe3}
\end{figure}

\begin{figure}[htbp] 
    \centering
    \includegraphics[width=0.7\textwidth]{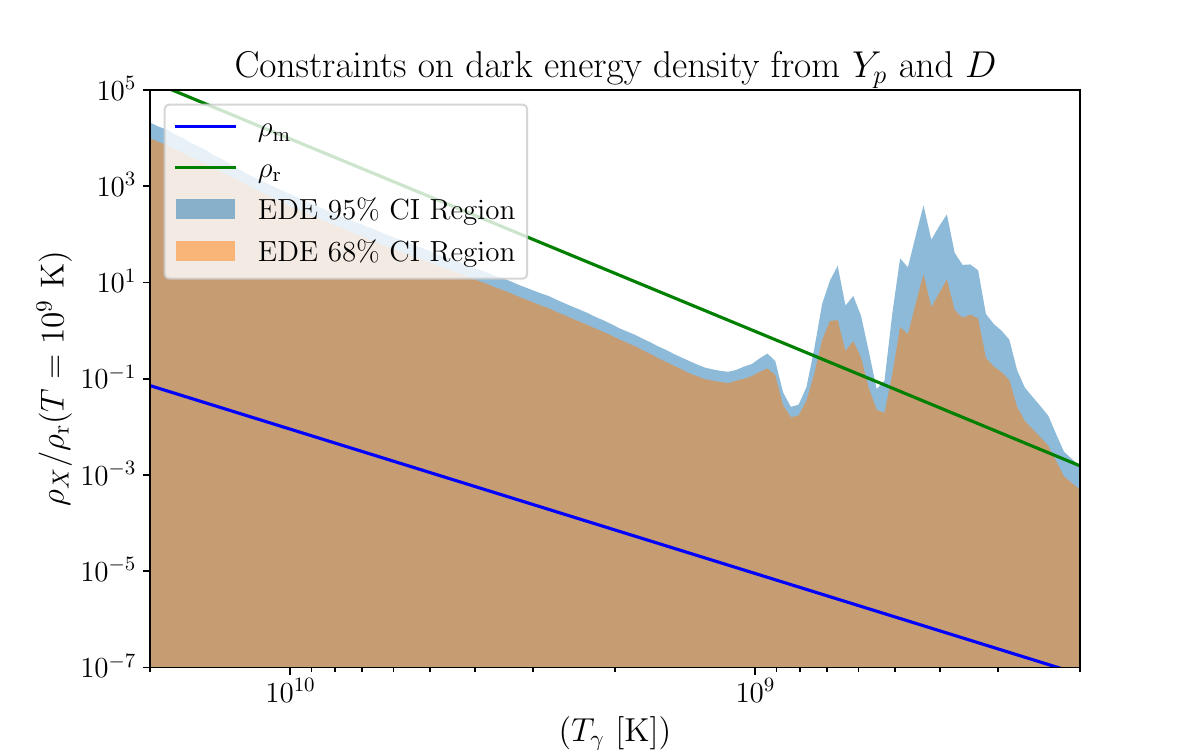} 
    \caption{Constraints on EDE from observed values of $Y_{p}$ and $D$, with a wide flat prior on $\eta_{10}\in[1,20]$. Shown are the matter and radiation densities, together with the 68\% and 95\% confidence intervals for EDE.}
    \label{fig:NoFlat}
\end{figure}

The analysis presented here also allows us to determine how future data might impact our ability to constrain the early expansion history. Figure~\ref{fig:NoHe3} presents the corresponding histories obtained after including a forecasted $1\%$ constraint on the helium-3 abundance. The addition of this prior produces minimal qualitative change in the allowed EDE behavior. 
As before, EDE is required to remain largely subdominant, with the same narrow windows that are more weakly constrained between $3\times10^{8}$~K and $5\times10^{8}$~K and  near or following the deuterium bottleneck.

In contrast, Figure~\ref{fig:NoFlat} illustrates the allowed EDE histories when only the helium mass fraction and deuterium abundance are constrained. At the same time, the prior on $\eta_{10}$ is relaxed to a flat range $[1,20]$. 
In this case,  constraints are much weaker throughout the entire BBN epoch,  allowing energy densities several orders of magnitude higher than in the previous scenarios.

However, this freedom is accompanied by extreme values of $\eta_{10}$ that are in strong tension with the precise constraints from CMB observations, highlighting the importance of an informed prior on the baryon-to-photon ratio.

Finally, when incorporating a prior on the lithium-7 abundance, no EDE models are found to be consistent within $2\sigma$ of the observed abundances. The resulting $\chi^2$ values from the output of \texttt{AlterBBN} fall outside the acceptable range, indicating that EDE alone is insufficient to resolve the lithium problem. This outcome suggests that additional physics beyond or alternative to EDE is required to explain the observed lithium-7 abundance.

\begin{figure}[h] 
    \centering
    \includegraphics[width=0.7\textwidth]{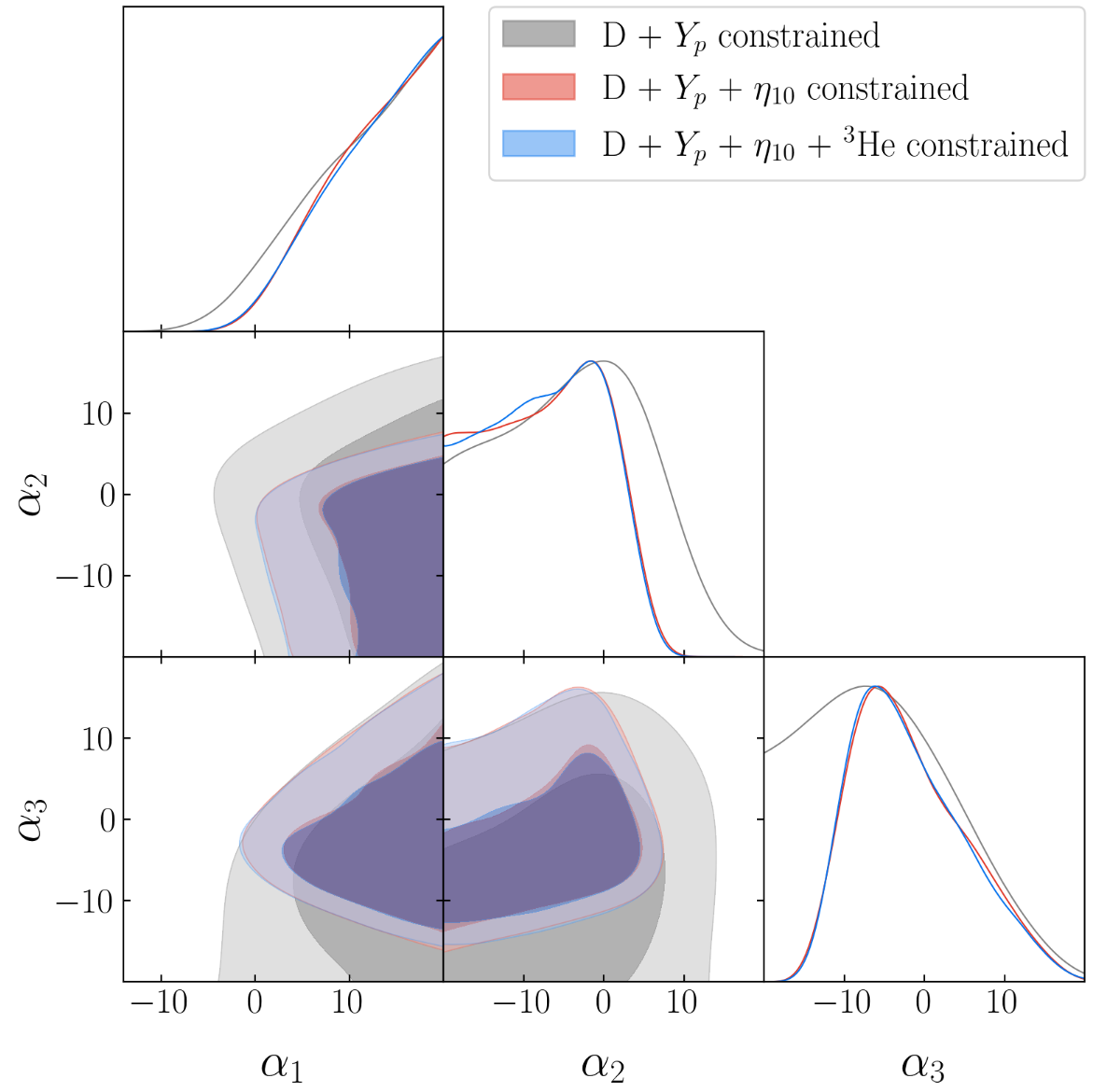}
    \caption{Constraints on the amplitudes of the principal components of the EDE density. As stated previously, there is not much additional constraining power provided by adding a 1\% constraint on the helium-3 abundance; however, we can see that including a constraint on $\eta_{10}$ derived from CMB observations makes the constraints on EDE derived from primordial abundance observations much more informative.
    }
    \label{fig:EigenstatesAmps}
\end{figure}

Overall, the combination of these scenarios illustrates that the constraints on EDE from BBN observables alone are able to limit additional energy density during the period ranging from weak freeze-out to the deuterium bottleneck and provide additional constraining power around the deuterium bottleneck and shortly afterwards. This behavior is expected, as the analysis relies solely on light element abundances.  These abundances are sensitive to the the neutron-to-proton ratio at the time that deuterium becomes bound, which depends on the ratio at weak freeze-out and the time elapsed between weak freeze-out and deuterium formation, both of which are sensitive to the expansion rate around this period.  The abundances are also sensitive to the expansion rate during and after the deuterium bottleneck, when the process of deuterium burning assembles the heavier nuclides. Figure~\ref{fig:EigenstatesAmps} presents the triangle plot from our MCMC analysis, showing constraints on the amplitudes of the principal components in the three scenarios discussed. This visualization highlights the differences in parameter degeneracies and allowed EDE amplitudes across the cases, providing a comprehensive summary of the constraints derived from light-element observations.

\section{Discussion and Conclusion}
\label{sec:Conclusion}

In this work, we have examined the constraints on early dark energy (EDE) during Big Bang Nucleosynthesis (BBN) using a combination of light element abundances and a principal component analysis (PCA) framework. By constructing a fiducial radiation-like EDE model and perturbing it in temperature bins, we generated a basis for a Fisher matrix analysis, identifying the eigenmodes to which BBN observables are most sensitive. These eigenmodes were then explored with a Markov Chain Monte Carlo (MCMC) approach, allowing for a model-independent reconstruction of the allowed EDE histories.

Our results indicate that, when constrained only by helium-4 and deuterium abundances, EDE is generally required to remain subdominant throughout most of the BBN epoch. 
We find a more weakly constrained region around $3\times10^8$~K to $5\times10^8$~K, and modest  EDE contributions are permitted near or just after the deuterium bottleneck.  However, we note that physically consistent models of EDE are unlikely to allow for spikes of energy density that would saturate the upper bounds in these narrow temperature ranges. 

The inclusion of a forecasted 1\% helium-3 prior produces minimal change in these constraints, implying that at this level of precision, observations of the primordial helium-3 abundance would not provide much new insight into the early expansion history. 
When the prior on the baryon-to-photon ratio $\eta_{10}$ is relaxed,  significantly larger EDE contributions are allowed throughout BBN; however, the resulting best-fit $\eta_{10}$ values in this scenario are in extreme tension with CMB measurements, emphasizing the importance of complementary cosmological data in constraining EDE.

Notably, the lithium-7 abundance remains incompatible with any allowed EDE histories, even within 2$\sigma$ of observational uncertainties. This indicates that EDE alone cannot resolve the longstanding lithium problem, and additional physics, whether new particle species, nonstandard interactions, or astrophysical processes, is likely required.

The triangle plot of eigenmode amplitudes from our MCMC analysis confirms that the strongest constraints on the expansion history arise between weak freeze-out and the deuterium bottleneck while also providing constraints on the evolution shortly after the deuterium bottleneck, consistent with the expectation that light element abundances are sensitive to the conditions during these periods. The analysis demonstrates that, while BBN provides a valuable early-universe probe of EDE, its constraining power is fundamentally limited to a specific temperature window, reinforcing the need to combine BBN constraints with other cosmological observables, such as the CMB and large-scale structure surveys, for a more complete picture.

Future work could extend this approach by incorporating updated lithium-7 measurements, exploring alternative dark energy parameterizations, or jointly analyzing BBN with CMB and large-scale structure data. Such studies would clarify whether EDE can play a significant role in early cosmology without conflicting with established observational constraints, and help determine what additional physics may be required to fully resolve discrepancies like the lithium problem.

\section*{Acknowledgments}
The authors would like to thank Bohua Li, Joshua Perez, Ana Segovia, and Paul Shapiro for helpful discussions. 
This work was supported by the US~Department of Energy under Grant~\mbox{DE-SC0010129}, by NASA through Grant~\mbox{80NSSC24K0665}, and by NSF through grant \mbox{AST-2510926}.
This work benefited from support provided by the Hamilton Undergraduate Research Scholars Program at SMU.
Computational resources for this research were provided by SMU’s O’Donnell Data Science and Research Computing Institute.

\bibliographystyle{utphys}
\bibliography{bbn}

\end{document}